\documentclass[letterpaper]{article} 

\ifdefined\aaaianonymous
    \usepackage[submission]{aaai2026}  
\else
    \usepackage{aaai2026}              
\fi

\usepackage{times}  
\usepackage{helvet}  
\usepackage{courier}  
\usepackage[hyphens]{url}  
\usepackage{graphicx} 
\usepackage{natbib}  
\usepackage{caption} 
\usepackage{amsmath}
\usepackage{amssymb}
\ifdefined\aaaianonymous
    \title{Speech Meets ELF: Audio Conditional Continuous-Target Diffusion\\for Speech Recognition and Translation}
\else
    \title{Speech Meets ELF: Audio Conditional Continuous-Target Diffusion\\for Speech Recognition and Translation}
\fi

\author{
    Xuanchen Li\equalcontrib\textsuperscript{\rm 1},
    Tianrui Wang\equalcontrib\textsuperscript{\rm 1},
    Yuheng Lu\textsuperscript{\rm 1},
    Zikang Huang\textsuperscript{\rm 1},
    Yu Jiang\textsuperscript{\rm 1},
    Chenghan Lin\textsuperscript{\rm 1},\\
    Chenrui Cui\textsuperscript{\rm 1},
    Ziyang Ma\textsuperscript{\rm 2},
    Xingyu Ma\textsuperscript{\rm 3},
    Chunyu Qiang\textsuperscript{\rm 1},
    Guochen Yu\textsuperscript{\rm 3},\\
    Xie Chen\textsuperscript{\rm 2},
    Longbiao Wang\thanks{Corresponding author.}\textsuperscript{\rm 1},
    Jianwu Dang\textsuperscript{\rm 1}
}
\affiliations{
    \textsuperscript{\rm 1}Tianjin University \ \ \ 
    \textsuperscript{\rm 2}Shanghai Jiao Tong University \ \ \ 
    \textsuperscript{\rm 3}Zhipu AI\\
}

\nocopyright

\begin{document}

\maketitle

\begin{abstract}
Speech-to-text (S2T) systems for recognition (ASR) and translation (S2TT) typically generate discrete text tokens.
In contrast, continuous-target language modelling performs generation in a continuous space, yet its potential for S2T remains unexplored.
To bridge this gap, we propose ELF-S2T, an audio-conditioned continuous-target generative model for S2T.
Built upon the pre-trained Embedded Language Flows (ELF) backbone, ELF-S2T processes speech via a frozen Whisper encoder and a single linear projector, prepending the resulting audio condition to the noisy text latent for in-context, flow-matching denoising.
To prevent the model from over-relying on its pre-trained text context, we introduce audio forcing during training, and further amplify the audio condition via classifier-free guidance at inference. Experiments on LibriSpeech and CoVoST2 show that ELF-S2T achieves competitive ASR and S2TT performance. Crucially, our error analysis reveals that, although ASR and S2TT errors look very different on the surface, both stem from the same underlying cause, a close distance confusion in the continuous latent space. This finding naturally aligns with the continuous representation generation paradigm, indicating a common semantic mapping process beneath recognition and translation.
Our code and pretrained models are publicly available at \url{https://github.com/Sslnon/ELF-S2T}.
\end{abstract}

\section{Introduction}
\label{sec:intro}

\begin{figure}[t]
\centering
\includegraphics[width=\columnwidth]{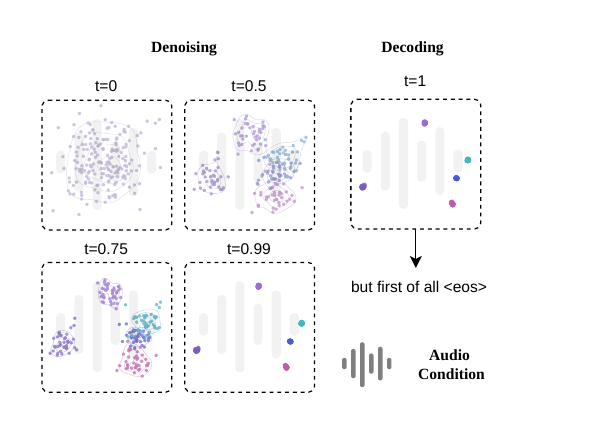}
\caption{ELF-S2T casts speech-to-text as audio-conditioned generation in a continuous text space. Starting from Gaussian noise at $t\!=\!0$, the text latent is denoised toward the target under the audio condition, and tokens are unembedded only at the final step $t\!=\!1$.}
\label{fig:teaser}
\end{figure}

Speech-to-text (S2T), comprising automatic speech recognition (ASR) and speech-to-text translation (S2TT), is dominated by discrete-token autoregressive backbones built on large language models.
Systems such as Whisper~\cite{radford2022robustspeechrecognitionlargescale}, SeamlessM4T~\cite{communication2023seamlessm4tmassivelymultilingual}, and Qwen-Omni~\cite{xu2025qwen25omnitechnicalreport,xu2025qwen3omnitechnicalreport} cast S2T as next-token prediction over a discrete vocabulary, conditioned on a continuous speech representation produced by an acoustic encoder.
More recent work explores discrete-token diffusion as an alternative non-autoregressive sampler in the same target space: TransFusion~\cite{baas2022transfusiontranscribingspeechmultinomial} casts ASR as multinomial diffusion over text tokens, iteratively refining a fully-noised token sequence into the transcript.
Whisfusion~\cite{kwon2025whisfusionparallelasrdecoding} adopts a masked-diffusion formulation, predicting masked tokens in parallel over multiple denoising rounds.
Both proposals keep the target space discrete and report ASR results only.

Despite the diversity of decoders, these systems largely share the same modelling premise: speech-to-text generation is ultimately performed in a discrete token space.
In contrast, continuous-target diffusion language models, exemplified by ELF~\cite{hu2026elfembeddedlanguageflows} and Cola-DLM~\cite{guo2026continuouslatentdiffusionlanguage}, have shown that text generation can be performed within a continuous space rather than over discrete tokens.
Because speech signals are inherently continuous, mapping them directly to a continuous text representation space offers a theoretically elegant and natural generation paradigm~\cite{xu2024comparingdiscretecontinuousspace}, avoiding the fragmentation introduced by forced discretisation~\cite{wang2025speechdiscretetokenscontinuous}.
This formulation naturally supports classifier-free guidance~\cite{ho2022classifierfreediffusionguidance} and other diffusion-domain techniques, and has demonstrated competitive performance in text-only language modelling.
However, its potential for speech-to-text generation remains largely unexplored.
The core difficulty of this paradigm lies in cross-modal alignment~\cite{wu2025languageoverrulesrevealingtext}.
Because pre-trained continuous text models possess strong internal language model priors, introducing acoustic representations often leads the model to over-rely on pure text context, thereby ignoring the input speech features~\cite{leng2024cursemultimodalitiesevaluatinghallucinations}.
This raises a fundamental question: how can we deeply couple a continuous-target language model with speech signals and use speech information to genuinely guide text generation?

To this end, this paper explores continuous-target language modelling for speech-to-text generation and proposes \textbf{ELF-S2T}.
Concretely, ELF-S2T uses the output of a speech encoder as the audio condition and, initialised from a pre-trained ELF model \cite{hu2026elfembeddedlanguageflows}, trains a continuous-time flow-matching model to generate text within a continuous embedding space.
Rather than predicting discrete tokens at every decoding step, ELF-S2T keeps the generation process in the continuous text-latent space and maps the final representation to tokens only at the last step.
To mitigate the challenge of over-relying on text priors, we propose an audio forcing mechanism during the training phase.
By degrading the clarity of the text latent variables, this mechanism forces the model to utilise acoustic features for prediction.
Concurrently, we introduce condition dropout during training, which enables the model to use classifier-free guidance~\cite{ho2022classifierfreediffusionguidance} at the inference stage to further enhance its dependence on audio conditions and overall robustness.
These mechanisms ensure that the generation process is genuinely grounded in the acoustic input.
Furthermore, the continuous target space provides a novel lens for observing the generation deviations of the model.
Although ASR errors mostly appear as sub-word garbling on the surface and S2TT errors as sentence-level semantic drift, probing the latent space reveals that both tend to exhibit close-distance confusion within the continuous latent space.
This finding indicates that these two tasks rely on a common semantic mapping process at a fundamental level.
We make the following three contributions:
\begin{itemize}
\item We propose ELF-S2T, the first adaptation of continuous-target language modelling to speech, reaching $5.69$\% WER on LibriSpeech and $28.55$ BLEU on CoVoST2 de$\to$en, and the first diffusion-based S2T model to report S2TT results.
\item Targeting the condition dependence difficulty in speech generation, we introduce audio forcing and classifier-free guidance~\cite{ho2022classifierfreediffusionguidance} strategies, effectively enhancing the dependence of the model on speech signals and its generation robustness, and we demonstrate that this approach scales consistently with backbone capacity.
\item Using the continuous target space as an analytical lens, we reveal that ASR and S2TT errors, despite appearing vastly different on the surface, actually share a unified measurable cause in the latent space, indicating that both tasks rely on a common semantic mapping process at a fundamental level.
\end{itemize}

\begin{figure*}[t]
\centering
\includegraphics[width=0.87\textwidth]{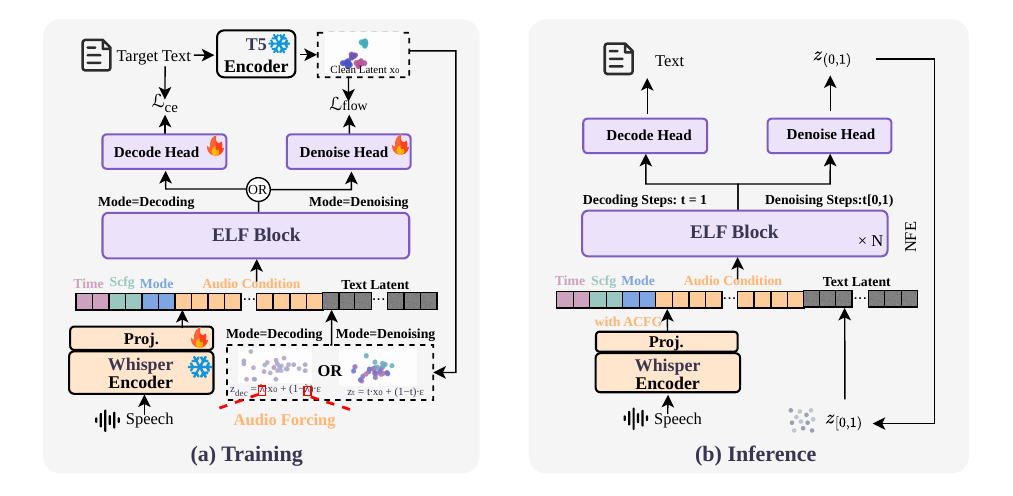}
\caption{Overview of ELF-S2T. A frozen Whisper encoder and a single projector turn speech into an audio condition that is prepended to the noisy text latent and denoised by the shared ELF backbone. \textbf{(a) Training.} A denoise head and a decode head are trained jointly, and our audio forcing weakens the text latent so the model must read the audio condition. \textbf{(b) Inference.} From Gaussian noise the backbone runs over SDE steps under audio guidance (ACFG), and tokens are decoded only at the final step.}
\label{fig:arch}
\end{figure*}

\section{Related Work}
\label{sec:related}

\paragraph{Discrete-token autoregressive speech-to-text generation.}
Most current ASR and S2TT systems follow a discrete generation paradigm, in which a continuous speech signal is first represented by an acoustic encoder and then decoded into text step by step.
Large-scale end-to-end systems such as Whisper~\cite{radford2022robustspeechrecognitionlargescale} and SeamlessM4T~\cite{communication2023seamlessm4tmassivelymultilingual} map speech directly to text.
In parallel, another line keeps a pre-trained text LLM and connects a speech encoder to the language model in order to perform speech-to-text generation~\cite{fathullah2023promptinglargelanguagemodels,yu2024connectingspeechencoder,ma2024embarrassinglysimple,tang2024salmonn,ma2025speechrecognitionmeets}.
Our work borrows this connection strategy at the architectural level, using a frozen Whisper encoder and connecting it to a generative text backbone through a single linear projector.

\paragraph{Discrete-token diffusion for speech-to-text.}

Beyond autoregressive decoding, recent work introduces diffusion-based non-autoregressive generation to speech-to-text.
This line no longer generates tokens from left to right, yet it still keeps discrete tokens as the modelling target, and it adapts discrete-state text diffusion methods such as D3PM~\cite{austin2021structureddenoising}, SEDD~\cite{lou2024discretediffusion}, and masked diffusion language models~\cite{sahoo2024simpleeffective} to speech.
For example, TransFusion~\cite{baas2022transfusiontranscribingspeechmultinomial} formulates ASR as multinomial diffusion over the token sequence, starting from a fully corrupted transcript and recovering the text through multiple iterative steps.
Whisfusion~\cite{kwon2025whisfusionparallelasrdecoding} instead uses a masked-diffusion decoder that recovers masked tokens in parallel across multiple prediction rounds.

These systems provide the closest points of comparison to ELF-S2T, because they also replace left-to-right autoregressive decoding with an iterative diffusion-style generation process. However, they differ from ELF-S2T in two key respects.
First, their denoising process still takes place in a discrete token space.
In contrast, ELF-S2T keeps the entire generation trajectory in a continuous embedding space and maps the continuous representation back to text only at the final step.
Second, existing discrete-token diffusion speech-to-text methods mainly report ASR results.
To our knowledge, no such system reports S2TT results systematically.

\paragraph{Continuous-target diffusion language models.}

Another key premise of our work comes from continuous-target language modelling, that text generation need not fall into a discrete token space at every step, but can instead unfold in a continuous representation space and commit to concrete tokens only at the final stage.
ELF~\cite{hu2026elfembeddedlanguageflows} is a representative method of this direction.
It defines the flow-matching process directly in the token-embedding space, so the model operates on continuous text representations throughout denoising and discretises only at the final step.
ELF-S2T inherits this formulation and further extends it to the audio-conditioned speech-to-text setting.
Related to ELF but following a different design path, Cola-DLM~\cite{guo2026continuouslatentdiffusionlanguage} also avoids modelling the full generation process directly in a discrete token space.
It obtains a compressed latent representation through a text VAE, and finally generates text with a separate conditional decoder.

Overall, ELF and Cola-DLM show that continuous-target modelling offers a path for text generation that differs from discrete-token decoding, and they have been competitive in text-only language modelling.
However, existing continuous-target diffusion language models have not been systematically studied under audio conditioning.
The question we therefore address is whether such a continuous-target text backbone can be effectively grounded on a speech signal and made to support speech-to-text tasks.

\section{Method}\label{sec:method}

As shown in Figure~\ref{fig:arch}, ELF-S2T performs speech-to-text as audio-conditioned generation in a continuous text space, prepending an audio condition to the ELF backbone and denoising from noise toward the latent of the target sentence, with tokens decoded only at the final step.
Speech recognition and translation use the same architecture and are trained as two separate models.

\subsection{Preliminaries}

Unlike discrete-token models that commit to a token at every position and every step, ELF generates text as a sequence of continuous embeddings.
A frozen T5 encoder~\cite{raffel2023exploringlimitstransferlearning} $\mathcal{E}$ maps a target sentence $y$ to a continuous latent $x_0=\mathcal{E}(y)\in\mathbb{R}^{T\times d}$, and a flow-matching transformer $f_\theta$ learns to denoise this latent from Gaussian noise.
Along a linear path between noise and data,
\begin{equation}
\label{eq:path}
z_t = t\,x_0 + (1-t)\,\epsilon, \qquad \epsilon\sim\mathcal{N}(0,I),
\end{equation}
with $t\in[0,1]$ where $t\!=\!1$ recovers the clean latent and $t\!=\!0$ is pure noise. The transformer is parametrised to output a prediction of the clean latent, $\hat{x}_0=f_\theta(z_t,t)$, and is trained by a flow-matching objective~\cite{lipman2023flowmatchinggenerativemodeling} that matches the velocity induced by this prediction to the target velocity of the path,
\begin{equation}
\label{eq:flow}
\begin{gathered}
\mathcal{L}_{\text{flow}} = \mathbb{E}_{t,\epsilon}\big[\,\|\hat{v}_t-v_t\|^2\,\big],\\
\hat{v}_t = \frac{f_\theta(z_t,t)-z_t}{1-t},\qquad
v_t = \frac{x_0-z_t}{1-t},
\end{gathered}
\end{equation}
where the predicted velocity $\hat{v}_t$ is induced by the clean-latent output $f_\theta(z_t,t)$, so this objective equals the clean-latent regression $\|f_\theta(z_t,t)-x_0\|^2$ reweighted by $1/(1-t)^2$.
A learned decode head $g_\phi$ then maps the final latent to tokens, $\hat{y}=\arg\max g_\phi(\hat{x}_0)$, the decode step in Figure~\ref{fig:arch} and the only point where the continuous trajectory is discretised.
ELF runs both $f_\theta$ and $g_\phi$ on one shared backbone and conditions it through a small set of learnable control tokens prepended to the sequence, carrying three signals, a timestep token for the diffusion time $t$, a self-conditioning token for the scale of ELF's self-conditioning pass, and a mode token that switches the backbone between two modes.
The mode token is the key signal for ELF, in denoise mode the denoise head $f_\theta$ predicts the clean latent, and in decode mode the decode head $g_\phi$ reads out tokens.
These two modes are the two branches in Figure~\ref{fig:arch}, and our audio forcing later acts on the decode mode, so we keep this control interface unchanged.
At inference, the model starts from Gaussian noise and a stochastic sampler iteratively integrates the flow to $t\!=\!1$, after which a single decode step produces the tokens.
We adopt this latent space, objective, and sampler as they are and initialise from a pre-trained ELF model, so this part is essentially unchanged from ELF.
Target length is handled by ELF's end-of-sequence padding scheme, and we truncate at the first end-of-sequence symbol at inference.

\subsection{Audio Conditioning}
As shown in Figure~\ref{fig:arch}, the audio condition comes from the encoder of a frozen Whisper-large-v3~\cite{radford2022robustspeechrecognitionlargescale}.
Let $h=\mathrm{Whisper}(s)\in\mathbb{R}^{M\times d_a}$ be the encoder output for speech input $s$.
A linear projector $P\in\mathbb{R}^{d_a\times d}$ maps it to the text-latent width, $a=hP\in\mathbb{R}^{M\times d}$.
We prepend this audio condition to the noisy text latent along the sequence axis and feed the concatenation to the ELF transformer, which now denoises conditioned on the audio,

\begin{equation}
\label{eq:cond}
[\,\tilde{a},\,\hat{x}_0\,] = f_\theta\big([\,a,\,z_t\,],\,t\big),
\end{equation}

attending jointly over the audio condition and the noisy text in one bidirectional stream.
We slice off the audio positions $\tilde{a}$ and keep only the text part $\hat{x}_0$.
The Whisper encoder stays frozen, the linear projector $P$ is the only newly added module, and the ELF backbone is fine-tuned from its pre-trained text weights.
Because the audio condition makes the sequence longer than ELF's pre-trained context, we extend its rotary position embeddings by interpolation, which keeps positions inside the pre-trained range and adds no parameters.

\subsection{Grounding Generation on Speech}

The central difficulty in moving ELF from text to speech is that a continuous-target text model can complete a sentence from its own text context, so a naive port lets the model lean on that context instead of listening to the audio.
We therefore shape the training and inference so that the audio condition becomes necessary rather than optional.

\paragraph{Audio forcing.}
The decode head $g_\phi$ is supervised on a separately noised latent
\begin{equation}
\label{eq:decode}
z_{\text{dec}} = \lambda\,x_0 + (1-\lambda)\,\epsilon,
\mathcal{L}_{\text{ce}} = \mathbb{E}\big[\mathrm{CE}\big(g_\phi(z^{\text{dec}}),\,y\big)\big],
\end{equation}
where $\lambda\in[0,1]$ is a random signal level that controls how much of the clean latent is retained.
ELF concentrates $\lambda$ near one, so the decode head sees an almost clean text latent that already reveals most of the answer, and the audio condition can be ignored.
We shift the distribution of $\lambda$ toward lower values, so the decode head is trained where the text latent alone is no longer sufficient and the model must read the audio condition to recover the content.
We also raise the share of training spent on the decode head, since speech-to-text targets are short and need a stronger decode signal to converge.
Audio forcing is the main driver of grounding in our recipe.

\paragraph{Audio guidance.}
To strengthen the conditioning at inference, we train an unconditional branch alongside the conditional one by dropping the entire audio condition on a small fraction of utterances.
At inference we combine the two branches by classifier-free guidance~\cite{ho2022classifierfreediffusionguidance},
\begin{equation}
o = w\,o_{\text{cond}} + (1-w)\,o_{\text{uncond}},
\end{equation}
where the scale $w$ of this audio classifier-free guidance (ACFG) controls how far the output is pushed toward the audio-conditioned prediction.
Setting $w$ to one recovers the conditional model, and a larger $w$ extrapolates away from the unconditional one.
We apply this combination at every denoising step and at the final decode step.

\begin{table*}[t]
\centering
\caption{Main results. WER is raw word error rate on LibriSpeech test-clean (lower is better), BLEU and chrF are on CoVoST2 de$\to$en test (higher is better). The Whisfusion and TransFusion numbers are taken from the original papers, all other rows are run through our pipeline. Decoder parameters count the text decoder only, the frozen Whisper encoder is excluded.}
\label{tab:main}
\footnotesize
\setlength{\tabcolsep}{6pt}
\begin{tabular}{llccc}
\hline
\textbf{Model} & \textbf{Decoder} & \textbf{Params} & \textbf{LS-clean} WER$\downarrow$ & \textbf{de$\to$en} BLEU\,/\,chrF$\uparrow$ \\
\hline
\multicolumn{5}{l}{\emph{Discrete-token autoregressive}} \\
Whisper-large-v3 (greedy) & Transformer & 907\,M & \textbf{1.97} & 26.23 / 54.38 \\
\hline
\multicolumn{5}{l}{\emph{Discrete-token diffusion}} \\
Whisfusion~\cite{kwon2025whisfusionparallelasrdecoding} & MDM & 301\,M & 8.30 & --- \\
TransFusion~\cite{baas2022transfusiontranscribingspeechmultinomial} & Multinomial & 253\,M & 6.10 & --- \\
\hline
\multicolumn{5}{l}{\emph{Continuous-target diffusion (ours)}} \\
ELF-S2T (ELF-L) & ELF-L & 653.4\,M & 5.69 & \textbf{28.55 / 54.91} \\
\hline
\end{tabular}
\end{table*}

\begin{table}[tb]
\centering
\caption{Audio forcing as a single-variable change on ELF-B, matched inference recipe (audio guidance $w\!=\!2.0$, $K\!=\!128$). WER on LibriSpeech test-clean.}
\label{tab:audioforce}
\footnotesize
\setlength{\tabcolsep}{8pt}
\begin{tabular}{lc}
\hline
\textbf{Setting} & \textbf{WER}$\downarrow$ \\
\hline
ELF default & 11.11 \\
\quad + audio forcing & \textbf{10.50} \\
\hline
\end{tabular}
\end{table}

\section{Experiments}\label{sec:exp}

\subsection{Setup}\label{sec:setup}

\paragraph{Data and tasks.}
We evaluate ELF-S2T on recognition and translation, training one model per task under the same architecture and recipe and evaluating each on its own test set.
The ASR model is trained on LibriSpeech 960\,h~\cite{Panayotov2015LibrispeechAA} and reported by word error rate on test-clean (2{,}620 utterances).
The S2TT model is trained on CoVoST2~\cite{wang2020covost2massivelymultilingual} German$\to$English (de$\to$en, 127\,k pairs) and reported by SacreBLEU and chrF on its test set (13{,}511 utterances).

\paragraph{Models.}
The conditioning encoder is a frozen Whisper-large-v3, and the continuous-target backbone is the ELF transformer in three sizes, ELF-B (105.9\,M), ELF-M (343.9\,M), and ELF-L (653.4\,M), each initialised from its public pre-trained\footnote{\url{https://github.com/lillian039/ELF}}.
The only newly trained parameters are the linear audio projector and the backbone weights, the Whisper encoder is never updated.

\paragraph{Training.}
All models are trained in bfloat16 with AdamW ($\beta_1\!=\!0.9$, $\beta_2\!=\!0.95$, no weight decay), a peak learning rate of $3\times10^{-4}$ with a $500$-step linear warmup followed by cosine decay, and gradient clipping at a global norm of $1.0$.
We use $8$ GPUs with a per-GPU batch of $4$ utterances, an effective batch of $32$, and train the recognition models for $50$k steps and the translation models for $100$k steps.
The recognition models are trained from the pre-trained ELF model, and each translation model is initialised from the trained recognition model of the same backbone.
Each step optimises one of two objectives chosen by a Bernoulli draw, the flow-matching objective on the path latent or the cross-entropy on the decode head, in a $1$/$1$ split.
For audio forcing we set the decode-head signal level well below the ELF default, centring its log-signal-to-noise schedule at $-0.5$ rather than $0.8$, so the decode head is trained where the text latent alone is uninformative.
To enable audio guidance, we drop the entire audio condition on $10$\% of utterances during training so that an unconditional branch is learned alongside the conditional one.

\paragraph{Inference.}
Unless stated otherwise we sample with the SDE sampler for $K\!=\!128$ steps with audio guidance $w\!=\!2.0$, and decode tokens once at the final step.

\paragraph{Metrics and baseline.}
ASR is scored by raw word error rate on LibriSpeech test-clean, and S2TT by SacreBLEU and chrF on the CoVoST2 de$\to$en test set.
As a discrete-token autoregressive reference, we run Whisper-large-v3 through the same pipeline with greedy decoding.
For discrete-token diffusion ASR we cite Whisfusion~\cite{kwon2025whisfusionparallelasrdecoding} and TransFusion~\cite{baas2022transfusiontranscribingspeechmultinomial} as reported, both trained on LibriSpeech-960 like ELF-S2T, since neither releases S2TT results.

\subsection{Main Results}\label{sec:main}
Table~\ref{tab:main} reports the main comparison.
On ASR, ELF-S2T with the ELF-L backbone reaches $5.69$\% WER on LibriSpeech test-clean.
This does not match the strongly supervised Whisper-large-v3 autoregressive reference ($1.97$\%), which marks a strong reference under the same pipeline, but it is lower than both discrete-token diffusion systems trained on LibriSpeech-960, Whisfusion ($8.3$\%) and TransFusion ($6.1$\%).
This shows that a continuous-target backbone, once grounded on speech, is competitive with discrete-token diffusion on recognition.

On S2TT, ELF-S2T reaches $28.55$ BLEU and $54.91$ chrF on CoVoST2 de$\to$en, above the Whisper-large-v3 reference ($26.23$ BLEU, $54.38$ chrF).
To our knowledge, this is the first diffusion-based speech-to-text system to report S2TT results, the prior discrete-token diffusion systems report recognition only.
The same architecture and recipe thus apply to both recognition and translation, each trained as its own model in a single continuous-target formulation.

\subsection{Ablation}\label{sec:abl}

We now isolate the effect of the grounding choices introduced in Method, using the ELF-B backbone for a fast and controlled comparison.

\paragraph{Audio forcing.}
Table~\ref{tab:audioforce} compares the ELF default decode-noise setting against audio forcing as a single-variable change, with everything else held fixed including the inference recipe.
Moving the unembed supervision to higher noise reduces ASR WER from $11.11$\% to $10.50$\% on LibriSpeech test-clean.
This behaviour follows from the unembed objective in Eq.~\ref{eq:decode}.
Under the ELF default the signal level $\lambda$ stays close to one, so the unembed is supervised on a nearly clean text latent that already encodes the target sentence.
The cross-entropy loss can then be minimised by reading the answer off this latent alone, the audio condition contributes little to lowering it, and the gradient therefore gives the model no incentive to attend to the speech.
The default schedule thus leaves open a shortcut that bypasses the audio.
Audio forcing removes this shortcut by shifting $\lambda$ toward lower values, where the text latent is too corrupted to recover the target on its own and the audio condition becomes the only reliable source of the content, which forces the model to ground its prediction on the speech.
This single change is the largest contributor to grounding in our recipe.

\begin{figure}[tb]
\centering
\includegraphics[width=\columnwidth]{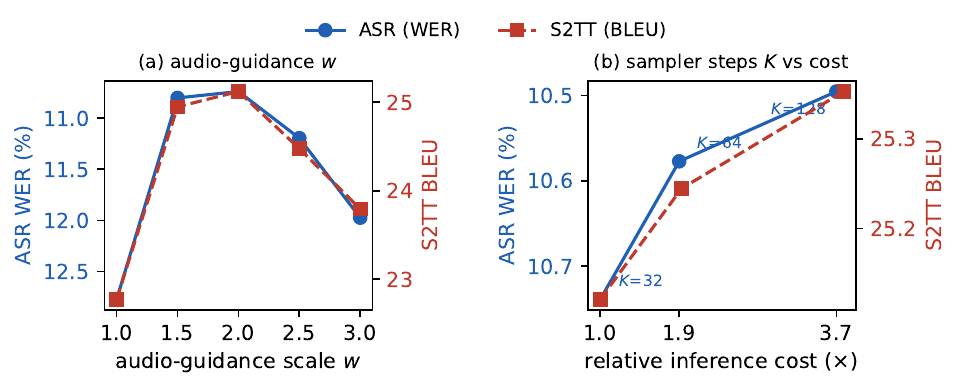}
\caption{Sweeps of the audio-guidance scale $w$ (a) and the sampler-step count $K$ (b) on ELF-B under the audio-forcing recipe, for ASR (blue, WER) and S2TT (red, BLEU). WER axes are inverted so that up is better on both curves. In (b), $K\!\in\!\{32,64,128\}$ is plotted against relative inference cost, normalised to $K\!=\!32$.}
\label{fig:cfg}
\end{figure}

\paragraph{Audio guidance and sampler steps.}
Figure~\ref{fig:cfg} sweeps the guidance scale $w$ and the number of sampler steps $K$.
Both tasks share a sweet spot near $w\!=\!2.0$, where ASR WER bottoms out and S2TT BLEU peaks, and performance degrades on either side, confirming that audio guidance is necessary but should not be pushed too far.
S2TT is the more sensitive of the two, consistent with its shorter and sharper target distribution.
More sampler steps help both tasks, and panel (b) plots this against the relative inference cost, which grows almost linearly with $K$ because audio guidance runs the backbone twice at every step.
The return is steeply diminishing, going from $K\!=\!32$ to $K\!=\!128$ costs about $3.7$ times the compute yet buys only $0.24$ WER and $0.23$ BLEU, with most of the gain already realised by $K\!=\!64$.

\subsection{Backbone Scaling}
\label{sec:scaling}

Table~\ref{tab:scaling} reports how performance scales with backbone capacity.
ASR WER falls from $10.50$\% at ELF-B to $5.69$\% at ELF-L, a $46$\% relative reduction, and S2TT BLEU rises from $25.35$ to $28.55$.
The continuous-target route is therefore not capped at small scale, and capacity is a consistent contributor that widens the margin over the discrete-token diffusion baselines.
On S2TT all three backbones train stably and BLEU increases monotonically with capacity, from $25.35$ at ELF-B through $27.31$ at ELF-M to $28.55$ at ELF-L.

\begin{table}[tb]
\centering
\caption{Backbone scaling under the full recipe. WER on LibriSpeech test-clean, BLEU on CoVoST2 de$\to$en test.}
\label{tab:scaling}
\footnotesize
\setlength{\tabcolsep}{6pt}
\begin{tabular}{lccc}
\hline
\textbf{Backbone} & \textbf{Params} & \textbf{WER}$\downarrow$ & \textbf{BLEU}$\uparrow$ \\
\hline
ELF-B & 105.9\,M & 10.50 & 25.35 \\
ELF-M & 343.9\,M & 7.61 & 27.31 \\
ELF-L & 653.4\,M & \textbf{5.69} & \textbf{28.55} \\
\hline
\end{tabular}
\end{table}

\section{Error Analysis}
\label{sec:error}

ELF-S2T exhibits error patterns that look different on the two tasks, recognition errors concentrate at the word level while translation errors manifest at the sentence level.
Examined only on the surface, the two appear to be distinct failure modes.
A closer reading already complicates this picture, both tasks contain a common family of semantic substitution in which the model emits a fluent but wrong word of related meaning, and the apparent difference is mainly one of granularity.
The continuous target space then lets us go further and examine where each error originates rather than only how it appears, and to test whether the two patterns share a common cause.
We first characterise the surface errors of each task, then probe the latent space in which generation occurs.
All analysis uses the ELF-B models for a fast and controlled comparison.

\subsection{Surface-Level Errors}

\paragraph{Recognition.}
On LibriSpeech test-clean, $29.0$\% of utterances are transcribed exactly, and the remaining errors are predominantly minor, two thirds of the imperfect utterances remain below $15$\% WER.
Substitutions account for the majority of edit operations, $4094$ substitutions against $971$ deletions and $452$ insertions, indicating that words are rarely omitted or inserted and are instead rendered incorrectly.
Within the substitutions, approximately $70$\% are form-level corruptions, in which a correct word is produced with an erroneous spelling or morphology, for example \emph{circumvention} rendered as \emph{circumcession} or \emph{recoiled} as \emph{recoild}, rather than a different valid word.
A second and smaller group of substitutions is different in kind.
Here the model emits a fully valid word that is acoustically and orthographically far from the reference, yet close to it in meaning, for example \emph{regardless} recognised as \emph{irrespective}, \emph{begun} as \emph{commenced}, or \emph{remembered} as \emph{recalled}, pairs whose character edit distance is $6$ to $10$ and cannot be explained by mishearing the surface form.
We call this group semantic substitution, and it is the more revealing of the two, since an exact synonym at edit distance $10$ can only arise if the model has committed to the meaning.

\paragraph{Translation.}
On CoVoST2 de$\to$en the output stays fluent and well-formed, output length is stable with $98.3$\% of hypotheses within half to one and a half times the reference length, and form-level corruption is negligible at $0.58$\% of hypotheses.
The errors are therefore not on the surface of words but in their selection.
The same semantic substitution seen in recognition is present here and is in fact common, the model keeps the sentence frame intact and replaces a single content word by an acoustically unrelated near synonym, for example \emph{safety} technology rendered as \emph{security} technology, or the \emph{recruitment} of troops as the \emph{promotion} of troops.
When such a swap drifts further from the reference it grows into a sentence-level drift, and at the extreme the output collapses into an unrelated sentence, so word-level semantic substitution and sentence-level drift are two ends of one continuum rather than separate modes.
Under sentence-BLEU, $22.1$\% of the test set is catastrophic (BLEU below $10$) and a further $47.3$\% is heavy (BLEU between $10$ and $30$), so on the surface the dominant failure is semantic drift rather than local corruption.

\paragraph{Comparison.}
The two tasks therefore share one error family and differ only in where it dominates.
Recognition is dominated by form-level corruption with semantic substitution as a clear minority, translation is dominated by semantic drift that ranges from a single swapped word to a fully rewritten sentence, with form-level corruption almost absent.
Table~\ref{tab:cases} gives representative examples of each.
Read this way the surface difference is one of granularity rather than of kind, a single substituted word in recognition and a phrase or clause in translation are the same act of choosing a wrong but meaning-related target.
This already hints that the two failures may share a cause, but surface form alone cannot establish it.
We examine the possibility directly in the latent space.

\begin{table*}[t]
\centering
\caption{Representative errors on the two tasks. Each task shows two error families, a form-level family that corrupts the surface of a word and a semantic-substitution family that replaces a word by an acoustically and orthographically unrelated word of related meaning. The semantic-substitution rows are the same phenomenon at two granularities, a single word in recognition and a phrase or clause in translation. Reference and hypothesis are shown as \emph{ref} and \emph{hyp}, with the changed span in bold.}
\label{tab:cases}
\footnotesize
\setlength{\tabcolsep}{6pt}
\begin{tabular}{p{0.11\textwidth}p{0.40\textwidth}p{0.40\textwidth}}
\hline
\textbf{Type} & \emph{ref} & \emph{hyp} \\
\hline
\multicolumn{3}{l}{\textbf{Recognition}} \\
spelling & must be done by \textbf{circumvention} & must be done by \textbf{circumcession} \\
morphology & returned to its place amidst the \textbf{tents} & returned to its place amidst the \textbf{tens} \\
semantic & remain divine \textbf{regardless} of opinion & remain divine \textbf{irrespective} of opinion \\
semantic & i have \textbf{begun} to love you & i have \textbf{commenced} to love you \\
\hline
\multicolumn{3}{l}{\textbf{Translation}} \\
semantic & the \textbf{safety} technology of the tunnel has been revised. & the \textbf{security} technology of the tunnel has been revised. \\
semantic & he began with the \textbf{recruitment} of new troops. & he began with the \textbf{promotion} of new troops. \\
drift & the receiver then gets a word that might be changed. & the recipient thus gets a possibly changed word. \\
collapse & it is marked in the dictionary. & go to du! \\
\hline
\end{tabular}
\end{table*}

\subsection{A Probe in the Latent Space}

Because ELF-S2T generates in a continuous text space, we can localise where a failure originates, at the audio-to-latent mapping that produces the prediction or at the latent-to-token unembed that converts it into text.
For each utterance we take the final latent $\hat{x}_0$ and compute its cosine similarity, after mean pooling over the sequence, to the frozen-encoder latents of the reference and of the model's own hypothesis, denoted $\cos_{\mathrm{ref}}$ and $\cos_{\mathrm{hyp}}$.
We compare a failure bucket against a normal bucket on each task, recognition garble against clean recognition and catastrophic translation against accurate translation.
As an upper bound we additionally report a ceiling, the cosine of a teacher-forced reconstruction that feeds the clean reference latent, which measures what the representation can attain under a perfect mapping.

\begin{table}[tb]
\centering
\caption{Encoder-space probe on ELF-B. Cosine of the final latent to the reference and to the model's own hypothesis, with a teacher-forced ceiling, averaged within each bucket. $\Delta=\cos_{\mathrm{hyp}}-\cos_{\mathrm{ref}}$, a positive value indicates that the latent is closer to the erroneous hypothesis than to the reference.}
\label{tab:probe}
\footnotesize
\setlength{\tabcolsep}{4.5pt}
\begin{tabular}{lrcccc}
\hline
\textbf{Bucket} & \textbf{n} & $\cos_{\mathrm{ref}}$ & $\cos_{\mathrm{hyp}}$ & \textbf{ceiling} & $\Delta$ \\
\hline
ASR normal & 50 & 0.619 & 0.620 & 0.945 & $+0.001$ \\
ASR garble & 100 & 0.545 & 0.656 & 0.949 & $+0.111$ \\
\hline
ST normal & 24 & 0.580 & 0.588 & 0.949 & $+0.008$ \\
ST catastrophic & 100 & 0.505 & 0.617 & 0.944 & $+0.112$ \\
\hline
\end{tabular}
\end{table}

If a failure were attributable to the unembed, the latent would lie near the reference, with $\cos_{\mathrm{ref}}$ high and comparable to the normal bucket, while the unembed nonetheless emitted incorrect tokens.
Table~\ref{tab:probe} shows the opposite.
On both tasks the failure buckets have a substantially lower $\cos_{\mathrm{ref}}$ than their normal counterparts, and the gap $\Delta$ increases from near zero to approximately $+0.11$, almost identically for recognition ($+0.111$) and translation ($+0.112$).
At failure, the latent has already drifted a short distance off the reference and toward a neighbouring but erroneous hypothesis before any token is unembedded.

The ceiling excludes the unembed as the cause.
Teacher-forced reconstruction attains approximately $0.945$ in every bucket, including the failures, so the representation and the unembed are jointly capable of expressing the correct answer.
What distinguishes success from failure is not this ceiling but the position of the sampled latent, which on failures is nudged a short distance toward a neighbouring, meaning-related point.
The unembed is therefore not responsible in either task, it faithfully renders a latent that the mapping step has already placed at a close but wrong neighbour of the target.

The surface analysis already shows a shared family of semantic substitution across the two tasks, differing mainly in granularity, and the probe now traces it to a single cause, the audio-conditioned mapping places the prediction at a close but wrong neighbour of the target in the latent space, a close-distance confusion that surfaces as one swapped word in recognition or, in translation, as a drift that can grow all the way to a fully rewritten sentence.
This deviation is geometrically mild in the latent space, the failed latent stays within the neighbourhood of the reference and sits only $\Delta\!\approx\!0.11$ closer to the erroneous hypothesis than to the reference on both tasks. The same small and almost identical offset shows up as only a single swapped word in recognition and as a catastrophic rewrite in translation, and this contrast, one geometric signature behind two very different surface severities, is precisely what unifies the error profiles of the two tasks.
The continuous target space makes this unification measurable, the close-distance confusion reads directly as a geometric distance in the latent space that we can quantify, and points to a common semantic mapping process beneath both tasks.
This also indicates a single direction for improvement, strengthening the audio-to-latent mapping, rather than separate remedies for two superficially distinct problems.

\section{Limitations}
\label{sec:limitations}

Our study has several limitations.
First, the recognition gap to the autoregressive reference ($5.69$\% versus $1.97$\% WER) is partly a metric artefact, WER charges the full edit cost of the morphology variants and meaning-preserving substitutions that our error analysis finds dominant, so raw WER likely understates the true quality of a model that commits to meaning before surface form.
Second, inference is iterative, $K$ steps with the backbone run twice per step under audio guidance, so it is more expensive than a single autoregressive pass.
Finally, we evaluate only English recognition and German$\to$English translation on read speech, leaving noisy speech and wider language pairs untested.

\section{Conclusion}
\label{sec:conclusion}

We presented ELF-S2T, the first adaptation of continuous-target language modelling to speech.
By conditioning a flow-matching backbone on a frozen speech encoder through a single linear projector, ELF-S2T performs recognition and translation entirely within a continuous text-embedding space, committing to discrete tokens only at the final step.
Trained as one model per task under the same architecture, ELF-S2T reaches $5.69$\% WER on LibriSpeech test-clean and $28.55$ BLEU on CoVoST2 de$\to$en, and to our knowledge it is the first diffusion-based speech-to-text model to report translation results.
Beyond the results, the continuous target space serves as an analytical lens on failure.
On the surface, recognition garbles sub-words and translation drifts at the sentence level, yet both are one semantic substitution that differs only in granularity.
A latent-space probe traces this to a single, directly measurable cause, the audio-conditioned mapping lands at a close but wrong neighbour of the target, so one mild and nearly identical geometry unifies surface costs that range from a single swapped word to a fully rewritten sentence.

\newpage

\bibliography{aaai2026}

\begin{thebibliography}{25}
\providecommand{\natexlab}[1]{#1}

\bibitem[{Austin et~al.(2021)Austin, Johnson, Ho, Tarlow, and van~den
  Berg}]{austin2021structureddenoising}
Austin, J.; Johnson, D.~D.; Ho, J.; Tarlow, D.; and van~den Berg, R. 2021.
\newblock Structured denoising diffusion models in discrete state-spaces.
\newblock In \emph{Proceedings of the 35th International Conference on Neural
  Information Processing Systems}, NIPS '21. Red Hook, NY, USA: Curran
  Associates Inc.
\newblock ISBN 9781713845393.

\bibitem[{Baas, Eloff, and
  Kamper(2022)}]{baas2022transfusiontranscribingspeechmultinomial}
Baas, M.; Eloff, K.; and Kamper, H. 2022.
\newblock TransFusion: Transcribing Speech with Multinomial Diffusion.
\newblock arXiv:2210.07677.

\bibitem[{Fathullah et~al.(2023)Fathullah, Wu, Lakomkin, Jia, Shangguan, Li,
  Guo, Xiong, Mahadeokar, Kalinli, Fuegen, and
  Seltzer}]{fathullah2023promptinglargelanguagemodels}
Fathullah, Y.; Wu, C.; Lakomkin, E.; Jia, J.; Shangguan, Y.; Li, K.; Guo, J.;
  Xiong, W.; Mahadeokar, J.; Kalinli, O.; Fuegen, C.; and Seltzer, M. 2023.
\newblock Prompting Large Language Models with Speech Recognition Abilities.
\newblock arXiv:2307.11795.

\bibitem[{Guo et~al.(2026)Guo, Zhao, Zhao, Nie, Zhu, Guo, Wang, Yang, Zhao,
  Wei, and Zeng}]{guo2026continuouslatentdiffusionlanguage}
Guo, H.; Zhao, Q.; Zhao, Y.; Nie, S.; Zhu, R.; Guo, Q.; Wang, F.; Yang, T.;
  Zhao, H.; Wei, G.; and Zeng, Y. 2026.
\newblock Continuous Latent Diffusion Language Model.
\newblock arXiv:2605.06548.

\bibitem[{Ho and Salimans(2022)}]{ho2022classifierfreediffusionguidance}
Ho, J.; and Salimans, T. 2022.
\newblock Classifier-Free Diffusion Guidance.
\newblock arXiv:2207.12598.

\bibitem[{Hu et~al.(2026)Hu, Qiu, Lu, Zhao, Li, Kim, Andreas, and
  He}]{hu2026elfembeddedlanguageflows}
Hu, K.; Qiu, L.; Lu, Y.; Zhao, H.; Li, T.; Kim, Y.; Andreas, J.; and He, K.
  2026.
\newblock ELF: Embedded Language Flows.
\newblock arXiv:2605.10938.

\bibitem[{Kwon et~al.(2025)Kwon, Ahn, Yun, Jwa, Choi, Park, Kim, Kim, Ryu, and
  Lee}]{kwon2025whisfusionparallelasrdecoding}
Kwon, T.; Ahn, J.; Yun, T.; Jwa, H.; Choi, Y.; Park, S.; Kim, N.-J.; Kim, J.;
  Ryu, H.~G.; and Lee, H.-J. 2025.
\newblock Whisfusion: Parallel ASR Decoding via a Diffusion Transformer.
\newblock arXiv:2508.07048.

\bibitem[{Leng et~al.(2024)Leng, Xing, Cheng, Zhou, Zhang, Li, Zhao, Lu, Miao,
  and Bing}]{leng2024cursemultimodalitiesevaluatinghallucinations}
Leng, S.; Xing, Y.; Cheng, Z.; Zhou, Y.; Zhang, H.; Li, X.; Zhao, D.; Lu, S.;
  Miao, C.; and Bing, L. 2024.
\newblock The Curse of Multi-Modalities: Evaluating Hallucinations of Large
  Multimodal Models across Language, Visual, and Audio.
\newblock arXiv:2410.12787.

\bibitem[{Lipman et~al.(2023)Lipman, Chen, Ben-Hamu, Nickel, and
  Le}]{lipman2023flowmatchinggenerativemodeling}
Lipman, Y.; Chen, R. T.~Q.; Ben-Hamu, H.; Nickel, M.; and Le, M. 2023.
\newblock Flow Matching for Generative Modeling.
\newblock arXiv:2210.02747.

\bibitem[{Lou, Meng, and Ermon(2024)}]{lou2024discretediffusion}
Lou, A.; Meng, C.; and Ermon, S. 2024.
\newblock Discrete diffusion modeling by estimating the ratios of the data
  distribution.
\newblock In \emph{Proceedings of the 41st International Conference on Machine
  Learning}, ICML'24. JMLR.org.

\bibitem[{Ma et~al.(2024)Ma, Yang, Yang, Gao, Wang, Du, Yu, Chen, Zheng, Zhang,
  and Chen}]{ma2024embarrassinglysimple}
Ma, Z.; Yang, G.; Yang, Y.; Gao, Z.; Wang, J.; Du, Z.; Yu, F.; Chen, Q.; Zheng,
  S.; Zhang, S.; and Chen, X. 2024.
\newblock An Embarrassingly Simple Approach for LLM with Strong ASR Capacity.
\newblock arXiv:2402.08846.

\bibitem[{Ma et~al.(2025)Ma, Yang, Yang, Gao, Wang, Du, Yu, Chen, Zheng, Zhang,
  and Chen}]{ma2025speechrecognitionmeets}
Ma, Z.; Yang, G.; Yang, Y.; Gao, Z.; Wang, J.; Du, Z.; Yu, F.; Chen, Q.; Zheng,
  S.; Zhang, S.; and Chen, X. 2025.
\newblock Speech recognition meets large language model: benchmarking, models,
  and exploration.
\newblock In \emph{Proceedings of the Thirty-Ninth AAAI Conference on
  Artificial Intelligence and Thirty-Seventh Conference on Innovative
  Applications of Artificial Intelligence and Fifteenth Symposium on
  Educational Advances in Artificial Intelligence}, AAAI'25/IAAI'25/EAAI'25.
  AAAI Press.
\newblock ISBN 978-1-57735-897-8.

\bibitem[{Panayotov et~al.(2015)Panayotov, Chen, Povey, and
  Khudanpur}]{Panayotov2015LibrispeechAA}
Panayotov, V.; Chen, G.; Povey, D.; and Khudanpur, S. 2015.
\newblock Librispeech: An ASR corpus based on public domain audio books.
\newblock \emph{2015 IEEE International Conference on Acoustics, Speech and
  Signal Processing (ICASSP)}, 5206--5210.

\bibitem[{Radford et~al.(2022)Radford, Kim, Xu, Brockman, McLeavey, and
  Sutskever}]{radford2022robustspeechrecognitionlargescale}
Radford, A.; Kim, J.~W.; Xu, T.; Brockman, G.; McLeavey, C.; and Sutskever, I.
  2022.
\newblock Robust Speech Recognition via Large-Scale Weak Supervision.
\newblock arXiv:2212.04356.

\bibitem[{Raffel et~al.(2023)Raffel, Shazeer, Roberts, Lee, Narang, Matena,
  Zhou, Li, and Liu}]{raffel2023exploringlimitstransferlearning}
Raffel, C.; Shazeer, N.; Roberts, A.; Lee, K.; Narang, S.; Matena, M.; Zhou,
  Y.; Li, W.; and Liu, P.~J. 2023.
\newblock Exploring the Limits of Transfer Learning with a Unified Text-to-Text
  Transformer.
\newblock arXiv:1910.10683.

\bibitem[{Sahoo et~al.(2024)Sahoo, Arriola, Schiff, Gokaslan, Marroquin, Chiu,
  Rush, and Kuleshov}]{sahoo2024simpleeffective}
Sahoo, S.~S.; Arriola, M.; Schiff, Y.; Gokaslan, A.; Marroquin, E.; Chiu,
  J.~T.; Rush, A.; and Kuleshov, V. 2024.
\newblock Simple and effective masked diffusion language models.
\newblock In \emph{Proceedings of the 38th International Conference on Neural
  Information Processing Systems}, NIPS '24. Red Hook, NY, USA: Curran
  Associates Inc.
\newblock ISBN 9798331314385.

\bibitem[{{Seamless Communication}, Barrault
  et~al.(2023)}]{communication2023seamlessm4tmassivelymultilingual}
{Seamless Communication}; Barrault, L.; et~al. 2023.
\newblock SeamlessM4T: Massively Multilingual \& Multimodal Machine
  Translation.
\newblock arXiv:2308.11596.

\bibitem[{Tang et~al.(2024)Tang, Yu, Sun, Chen, Tan, Li, Lu, Ma, and
  Zhang}]{tang2024salmonn}
Tang, C.; Yu, W.; Sun, G.; Chen, X.; Tan, T.; Li, W.; Lu, L.; Ma, Z.; and
  Zhang, C. 2024.
\newblock SALMONN: Towards Generic Hearing Abilities for Large Language Models.
\newblock In \emph{The Twelfth International Conference on Learning
  Representations}.

\bibitem[{Wang, Wu, and Pino(2020)}]{wang2020covost2massivelymultilingual}
Wang, C.; Wu, A.; and Pino, J. 2020.
\newblock CoVoST 2 and Massively Multilingual Speech-to-Text Translation.
\newblock arXiv:2007.10310.

\bibitem[{Wang et~al.(2025)Wang, Li, Cui, Yang, Chen, and
  Meng}]{wang2025speechdiscretetokenscontinuous}
Wang, D.; Li, J.; Cui, M.; Yang, D.; Chen, X.; and Meng, H. 2025.
\newblock Speech Discrete Tokens or Continuous Features? A Comparative Analysis
  for Spoken Language Understanding in SpeechLLMs.
\newblock arXiv:2508.17863.

\bibitem[{Wu et~al.(2025)Wu, Tang, Zheng, and
  Jiang}]{wu2025languageoverrulesrevealingtext}
Wu, H.; Tang, M.; Zheng, X.; and Jiang, H. 2025.
\newblock When Language Overrules: Revealing Text Dominance in Multimodal Large
  Language Models.
\newblock arXiv:2508.10552.

\bibitem[{Xu et~al.(2025{\natexlab{a}})Xu, Guo, He, Hu, He, Bai, Chen, Wang,
  Fan, Dang, Zhang, Wang, Chu, and Lin}]{xu2025qwen25omnitechnicalreport}
Xu, J.; Guo, Z.; He, J.; Hu, H.; He, T.; Bai, S.; Chen, K.; Wang, J.; Fan, Y.;
  Dang, K.; Zhang, B.; Wang, X.; Chu, Y.; and Lin, J. 2025{\natexlab{a}}.
\newblock Qwen2.5-Omni Technical Report.
\newblock arXiv:2503.20215.

\bibitem[{Xu et~al.(2025{\natexlab{b}})Xu, Guo, Hu, Chu, Wang, He, Wang, Shi,
  He, Zhu, Lv, Wang, Guo, Wang, Ma, Zhang, Zhang, Hao, Guo, Yang, Zhang, Ma,
  Wei, Bai, Chen, Liu, Wang, Yang, Liu, Ren, Zheng, Men, Zhou, Yu, Yang, Yu,
  Zhou, and Lin}]{xu2025qwen3omnitechnicalreport}
Xu, J.; Guo, Z.; Hu, H.; Chu, Y.; Wang, X.; He, J.; Wang, Y.; Shi, X.; He, T.;
  Zhu, X.; Lv, Y.; Wang, Y.; Guo, D.; Wang, H.; Ma, L.; Zhang, P.; Zhang, X.;
  Hao, H.; Guo, Z.; Yang, B.; Zhang, B.; Ma, Z.; Wei, X.; Bai, S.; Chen, K.;
  Liu, X.; Wang, P.; Yang, M.; Liu, D.; Ren, X.; Zheng, B.; Men, R.; Zhou, F.;
  Yu, B.; Yang, J.; Yu, L.; Zhou, J.; and Lin, J. 2025{\natexlab{b}}.
\newblock Qwen3-Omni Technical Report.
\newblock arXiv:2509.17765.

\bibitem[{Xu et~al.(2024)Xu, Zhang, Yu, Wu, and
  Yu}]{xu2024comparingdiscretecontinuousspace}
Xu, Y.; Zhang, S.-X.; Yu, J.; Wu, Z.; and Yu, D. 2024.
\newblock Comparing Discrete and Continuous Space LLMs for Speech Recognition.
\newblock arXiv:2409.00800.

\bibitem[{Yu et~al.(2023)Yu, Tang, Sun, Chen, Tan, Li, Lu, Ma, and
  Zhang}]{yu2024connectingspeechencoder}
Yu, W.; Tang, C.; Sun, G.; Chen, X.; Tan, T.; Li, W.; Lu, L.; Ma, Z.; and
  Zhang, C. 2023.
\newblock Connecting Speech Encoder and Large Language Model for ASR.
\newblock arXiv:2309.13963.

\end{thebibliography}

\end{document}